\documentclass[sigconf]{acmart}

\usepackage{seqsplit}
\let\oldtexttt\texttt
\renewcommand{\texttt}[1]{\oldtexttt{\seqsplit{#1}}}
\usepackage{tabularx}
\usepackage{xcolor}
\usepackage{tikz}
\usetikzlibrary{positioning, fit, arrows.meta, backgrounds}
\usepackage{xspace}
\newcommand{\sol}{{\textsc{Drishti}}\xspace}

\setcopyright{none}
\acmConference[CPSIoTSec~'26]{8th Joint Workshop on CPS \& IoT Security and Privacy}{2026}{The Hague, Netherlands}
\acmYear{2026}

\begin{document}

\title{\sol: AI-Led Human-Directed Vulnerability Auditing for 5G Cores}

\author{Sriram Ramachandran}
\orcid{0000-0003-2344-765X}
\email{sriramr@a-star.edu.sg}
\affiliation{%
  \institution{A*STAR Institute of Advanced Intelligence and Computing}
  \country{Singapore}}

\author{Levente Csikor}
\orcid{0000-0002-1837-2158}
\email{csikorl@a-star.edu.sg}
\affiliation{%
  \institution{A*STAR Institute of Advanced Intelligence and Computing}
  \country{Singapore}}

\author{Dinil Mon Divakaran}
\orcid{0000-0001-8706-432X}
\email{divakaran@a-star.edu.sg}
\affiliation{%
  \institution{A*STAR Institute of Advanced Intelligence and Computing}
  \country{Singapore}}

\begin{abstract}
Candidate generation for open-source vulnerabilities is no longer scarce.
AI-assisted code review now produces defect candidates cheaply, and industry programs pair them with expert human triage.
The remaining scarcity is validation and impact assessment, and the gap is largest in critical-infrastructure software like 5G cores. 
Here, validation has four costs: \emph{verification}, \emph{reachability}, \emph{impact}, and \emph{fix-completeness}.

We present \sol, an AI-led human-directed vulnerability audit framework with four components, one per cost: (i) an anti-pattern catalog for verification, (ii) critical-path triage for reachability, (iii) concentric validation for impact, and (iv) patch-review for fix-completeness.
Across audits of Open5GS and free5GC, \sol produced three findings. The first is a pre-authentication NULL-dereference in the Open5GS NRF multipart parser, fixed upstream with a CVE requested. The second is an ASN.1-PER memory amplification in the free5GC NGAP decoder. A 2-byte input from a rogue gNodeB OOM-kills the AMF in 6.2~seconds. The third is a defective patch on CVE-2025-69248 whose defense-in-depth check is dead code before authentication.

\end{abstract}



\maketitle

\section{Introduction}
\label{sec:introduction}
In April 2026, Google restructured the Chrome and Android Vulnerability Reward Programs to lower payouts on lower-complexity findings, citing that AI tools had made such bugs easier to discover~\cite{gvrp2026}. 
Two months later, Chrome 149 shipped 429 security fixes in a single stable release~\cite{chromereleases2026}. 
Both events point to the same shift: candidate generation is no longer scarce. 
The scarcity has moved to validation and impact assessment. 

AI-generated candidates are unreliable at both ends of the
vulnerability lifecycle. 
On CyberGym's $1,507$-instance benchmark across $188$ projects, frontier agents reproduce the target in only $17-22$\% of cases, surfacing $18$ incomplete patches across $15$ projects~\cite{wang2025cybergym}.
Repair is subtler: patches from frontier agents pass their own test suites at indistinguishable rates, but expert review finds many ``successful'' fixes only guard the crash site or mask corrupted state, leaving the defect intact~\cite{wang2026kumushi}. 
The metrics that signal whether a candidate is real, or a patch actually fixes it, no longer discriminate.

This gap widens in domains where validation itself is expensive. 
A 5G core is one such domain.
Rather than a single application, it is a set of coordinating software components, i.e., network functions, that together handle authentication, session management, and service discovery for connected devices.
Its validation work has four costs. \emph{Verification} determines whether a candidate is a real defect. \emph{Reachability} determines whether a realistic attacker can trigger it. \emph{Impact} measures the operator-visible consequence. \emph{Fix-completeness} checks whether a patch covers the vulnerability class rather than the reported instance.
The last three take on a distinctly 5G-specific character. 
Reachability requires a full testbed. 
A researcher must emulate a Radio Access Network (RAN), a User Equipment (UE), and the core itself. 
Expertise in 3GPP protocols is required to construct inputs an attacker could actually deliver.
Impact depends on which procedures carry operator-visible consequence, specified across a tightly interlocking set of 3GPP documents. 
The 3GPP corpus comprises over 13{,}000 technical specifications~\cite{karim2023spec5g} organized into cross-referencing series (architecture, protocol, security, radio), and 5G spans successive releases whose specifications continue to change through quarterly change requests.
Achieving fix-completeness is complicated by two main factors: i)~the retention of 4G legacy wire formats, 
(namely NAS and NGAP protocols encoded via ASN.1 Packed Encoding Rules)
and ii)~under-the-hood deployment configurations (e.g., IPsec on the N2 interface) that cannot be detected through source code analysis alone.
Unlike traditional web infrastructure, where any user can execute attacker-controlled code by visiting an attacker-controlled URL, an end-user device cannot directly target a 5G network function in this manner.
Consequently, the consumer-scale validation loop that drove Chrome's routine 429-fix patching cycle is entirely absent in cellular ecosystems.
Despite this lack of consumer-driven fuzzing, these open-source cores are increasingly used in production and critical-infrastructure deployments. Their cellular networks carry emergency, industrial, or utility traffic.
A core outage can therefore prevent devices from establishing new sessions. Section~\ref{sec:case-nrf} reports the measured control-plane cascade and PFCP loss.
For example, Open5GS has been field-validated against a commercial Ericsson RAN, achieving more than 200~Mbps per UE in 90\% of field measurements across a deployment spanning a 30-km diameter~\cite{ara2025realworld}.

In this work, we present \sol\footnote{\sol originates from Sanskrit and means observe, focused gaze, insight, etc.}, an \emph{AI-led human-directed} vulnerability audit framework for open-source 5G cores.
The pairing of frontier-model candidate generation with expert human triage is now industry practice~\cite{openai2026patch}. \sol specializes that pattern for the cellular domain.
Its four components address the four costs identified above.
(i)~A domain-specific anti-pattern catalog for \emph{verification} by narrowing candidates to structural forms known to recur in cellular-core codebases. 
(ii)~A critical-path triage for \emph{reachability} analysis by partitioning candidates by attacker-reachable procedure. 
(iii)~Concentric validation tests \emph{impact} at increasing levels of deployment scope and cost.
(iv)~Patch-review tests \emph{fix-completeness} by treating fix coverage as a first-class audit artifact rather than a downstream question.
Applied to Open5GS and free5GC, \sol uncovered three defects. 
The first is a NULL-dereference in the multipart parser of Open5GS's NRF (Network
Repository Function), fixed upstream and submitted
to MITRE for a CVE.
The second is an ASN.1-PER memory amplification in free5GC's NGAP decoder: a 2-byte malformed pre-authentication input from a rogue gNodeB causes the AMF to be OOM-killed in 6.2~seconds.
The third is a defective patch for CVE-2025-69248 in free5GC's NAS
layer, where the shipped fix closes the original AMF panic but leaves
the intended defense-in-depth validation block unreachable through two
coding errors, confirmed by unit tests against the released patch (Appendix~\ref{app:patch-details}).

\section{Background and Related Work}
\label{sec:related_work}

\subsection{5G Core Architecture}
\label{sec:5gc-primer}
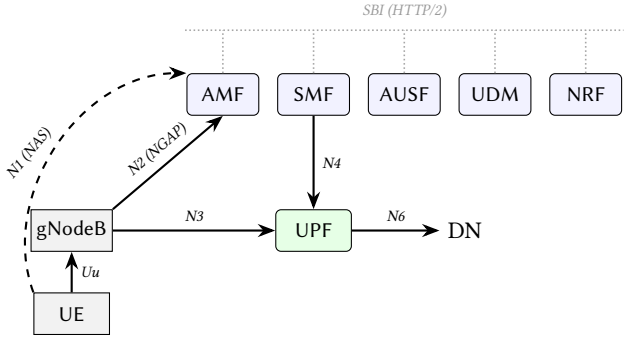
\begin{figure}[t]
\centering
\begin{tikzpicture}[
  nf/.style={draw, rounded corners=2pt, minimum width=0.95cm,
             minimum height=0.55cm, font=\small\sffamily,
             inner sep=2pt, fill=blue!5},
  actor/.style={draw, minimum width=1.0cm, minimum height=0.55cm,
                font=\small\sffamily, inner sep=2pt, fill=gray!10},
  up/.style={draw, rounded corners=2pt, minimum width=1.0cm,
             minimum height=0.55cm, font=\small\sffamily,
             inner sep=2pt, fill=green!10},
  ifc/.style={font=\scriptsize\itshape},
  sbi/.style={densely dotted, semithick, gray!70},
  arr/.style={->, >=Stealth, thick},
]

\node[nf] (amf)  at (2.5, 2.0) {AMF};
\node[nf] (smf)  at (3.7, 2.0) {SMF};
\node[nf] (ausf) at (4.9, 2.0) {AUSF};
\node[nf] (udm)  at (6.1, 2.0) {UDM};
\node[nf] (nrf)  at (7.3, 2.0) {NRF};

\draw[sbi] (2.0, 2.85) -- (7.8, 2.85);
\node[font=\scriptsize\itshape, gray!80] at (4.9, 3.10) {SBI (HTTP/2)};
\foreach \n in {amf, smf, ausf, udm, nrf}
  \draw[sbi] (\n.north) -- (\n.north |- 0, 2.85);

\node[actor] (gnb) at (0.5, 0.2)  {gNodeB};
\node[actor] (ue)  at (0.5, -0.9) {UE};

\node[up] (upf) at (3.7, 0.2) {UPF};
\node     (dn)  at (5.7, 0.2) {DN};

\draw[arr] (ue) -- node[ifc, right]{Uu} (gnb);
\draw[arr] (gnb.north east) -- node[ifc, above, sloped]{N2 (NGAP)} (amf.south);
\draw[arr] (gnb) -- node[ifc, above]{N3} (upf);
\draw[arr] (upf) -- node[ifc, above]{N6} (dn);
\draw[arr] (smf.south) -- node[ifc, right]{N4} (upf.north);
\draw[arr, dashed] (ue.north west) to[bend left=55]
    node[ifc, above left, sloped, pos=0.55]{N1 (NAS)} (amf.north west);

\end{tikzpicture}
\caption{5G core architecture. Five control-plane network functions
(AMF, SMF, AUSF, UDM, NRF) communicate over the HTTP/2-based
Service-Based Interface (SBI). Each function connects to the bus
above. NAS signalling flows logically between the UE and the AMF
over N1 (dashed). NGAP flows between the gNodeB and the AMF over
N2. User traffic passes through the UPF via N3 and out to the data network
(DN) via~N6.}
\Description{Diagram of the 5G core architecture showing control-plane and user-plane network functions and their connections.}
\label{fig:5gc-arch}
\end{figure}

The 3GPP 5G core is organized as a service-based architecture (Fig.~\ref{fig:5gc-arch}). Control-plane network functions expose services to each other over an HTTP/2-based Service-Based Interface (SBI), while the User Plane Function (UPF) forwards subscriber traffic. 
Five control-plane functions are relevant to our work here.  
The Access and Mobility Management Function (AMF) handles device registration and mobility. 
The Session Management Function (SMF) controls user sessions. 
The Authentication Server Function (AUSF) authenticates subscribers, while the Unified Data Management (UDM) function stores subscription data.
The Network Repository Function (NRF) registers functions and serves discovery queries.

Three protocol surfaces expose the core to attacker-influenced input. 
Non-Access Stratum (NAS) signalling carries authentication and registration messages between the UE and the AMF, encoded as compact type-length-value (TLV) fields. 
NGAP carries messages between the RAN and the AMF over the N2
interface, encoded in ASN.1 Packed Encoding Rules (PER), inherited from 4G LTE. 
SBI carries inter-function requests as HTTP/2 with JSON or multipart payloads. 
N2 IPsec is optional and often disabled in lab deployments, leaving NGAP effectively cleartext to radio-proximate attackers.

A device registration exemplifies the interaction: the UE sends a
NAS Registration Request over NGAP to the AMF, which invokes AUSF and UDM services over SBI to authenticate the subscriber and then registers session state with the NRF. 
Each hop is a potential audit surface. 
Independent open-source implementations of the 5G core include Open5GS (C), free5GC (Go), and Ella Core (Go, free5GC-derived and released for production private-network deployments in January 2026~\cite{ellanetworks2026ellacore}). 
This paper audits the two upstream projects - Open5GS and free5GC. Since Ella Core reuses free5GC's protocol libraries, library-level findings on free5GC carry over to it.

\subsection{5G Security Testing}
\label{sec:5g-testing}
Cellular networks are critical infrastructure, and cellular-security testing has a foundational lineage.
LTEInspector~\cite{hussain2018lteinspector} combined model checking with cryptographic protocol verification to find ten attacks on the 4G NAS layer under a Dolev-Yao adversary, validating eight in an SDR testbed. 
5Greplay~\cite{salazar20215greplay} introduced traffic-level fuzzing for 5G control protocols, injecting modified and replayed traffic into core services such as the AMF, and reported that the target services accept the altered traffic---an early instance of the AMF-input fragility class this paper revisits.

Three current-generation systems test open-source 5G core implementations directly. 
CoreCrisis~\cite{dong2025corecrisis} learns a finite-state model of the NAS/N1 registration surface from black-box benign runs and drives stateful mutation fuzzing against the learned FSM, targeting the same pre-authentication surface as this paper's parser finding through a different strategy.
FivGeeFuzz~\cite{chen2025cross} takes the opposite direction: it fuzzes SBIs post-authentication using grammar templates derived from 3GPP OpenAPI specifications, and discovered a scope-check bypass in free5GC's OAuth handling. 
Kairos~\cite{guo2026kairos} perturbs execution timing at network functions rather than input content, surfacing 20 new vulnerabilities across four cores including 11 CVE-assigned bugs---an orthogonal defect class.

On the radio side, 5Ghoul~\cite{garbelini20255ghoul} disclosed 14 pre-authentication vulnerabilities in Qualcomm and MediaTek 5G basebands via rogue-gNB downlink injection, and SNI5GECT~\cite{luo2025sni5gect} achieved the same class of attack without a rogue gNB using real-time sniffing and downlink injection over srsRAN. 

Together, these works establish an active 5G-testing landscape. 
\sol differs from every entry in three ways. First, it reads source instead of fuzzing the wire, the FSM, or timing. Second, it applies an anti-pattern catalog for cellular-core code. Third, it audits shipped patches as first-class artifacts. The cited systems do not report shipped-patch audits as outputs.

\subsection{Protocol-Specification Analysis with LLMs}
\label{sec:spec-nlp}
A separate line of research asks whether LLMs can extract structure directly from protocol specifications. 
SPEC5G~\cite{karim2023spec5g} introduced the first large-scale NLP dataset for 5G protocols---3.5~million sentences drawn from more than 13k cellular-network specifications and 13 online websites---and showed that BERT-family models pretrained on it improve downstream classification and summarization of security-relevant spec text.
PSMBench~\cite{shen2025psmbench} extends the direction to multi-protocol state-machine extraction across 14 protocols (including TCP, BGP-4, SIP, and DHCPv4), spanning $1,580$ pages of RFCs and evaluated on nine LLMs including GPT-4o-mini and Claude~3. 
Its central finding is a persistent state-transition gap: the best-performing model identifies states with F1-score $0.72$ but recovers transitions with F1-score $\leq 0.38$, meaning long-range dependency tracking and event-action disambiguation remain unreliable when models read specifications
raw. 

\sol takes the inverse approach. 
Rather than asking LLMs to extract protocol semantics from specifications, we encode those semantics once in the anti-pattern catalog and 3GPP-driven triage, then apply them during code review.

\subsection{AI-based vulnerability discovery}
\label{sec:ai-vuln}

Industry and governments have committed to AI-assisted vulnerability discovery at scale. 
OpenAI's \textit{Patch the Planet}~\cite{openai2026patch}, launched in June 2026, deploys frontier models against 19 major open-source projects (including Linux, OpenBSD, cURL, and Python).
The initiative embeds dedicated security engineers to review every AI-surfaced finding before it reaches maintainers, framing itself explicitly around the principle that ``discovery alone does not protect users.'' 
The program has identified hundreds of security issues and merged dozens of patches.
DARPA's two-year AI Cyber Challenge (AIxCC,~\cite{darpa2025aixcc}), concluded August 2025, demonstrated the same pattern at competition scale. 
Cyber reasoning systems from seven teams analyzed 54 million lines of code, discovered 86\% of synthetic vulnerabilities across 63 challenges, and patched 68\% of those found---at an average cost of \$152 per task. 
Together, these investments validate a common pattern: AI-first candidate generation paired with expert human triage, disclosure, and patch-review.

While industry programs validate the pattern operationally, academic systems investigate the design question that follows: how to structure LLM invocation so its outputs are review-worthy. 
VULSOLVER~\cite{li2025vulsolver} frames vulnerability detection as constraint satisfaction. For each candidate path, an LLM checks whether reachability and sink-state trigger constraints hold simultaneously. Static analysis supplies context for each caller and callee.
Its ablation shows that context maintenance across subtasks, not model capability alone, is what carries the approach---yielding 100\% recall on the OWASP benchmark and 15 zero-days across real projects. 
Argus~\cite{liang2026argus} takes a broader multi-agent stance: LLM agents drive dataflow reasoning on top of CodeQL through a Retrieval-Recursion-Review loop, 
augmented by retrieval over public vulnerability databases (NVD, OSV, GHSA,
Snyk). 
The framework successfully identified several CVE-assigned zero-days across seven mature Java codebases.
Beyond individual systems, JITVUL~\cite{JitVul2025benchmarking} released a benchmark for repository-level vulnerability detection based on just-in-time commit analysis, providing a more realistic evaluation setting than function-in-isolation datasets.

Repair systems exhibit a parallel evolution. 
Kumushi~\cite{wang2026kumushi} combines diversified dynamic fault localization, i.e., crash-stack tracing, fuzzer-generated variant traces, and CodeQL dataflow, with evidence-weighted function-of-interest ranking, narrowing the LLM's focus pre-generation. 
On 178 C/C++ vulnerabilities across 30 projects, Kumushi produces plausible patches at rates statistically indistinguishable from Codex under three replay-based oracles, yet expert reviewers prefer Kumushi's patches 63.6\% of the time ($p=0.0065$). 
Kumushi traces the gap to nine symptom-fix strategies and argues that oracle-based evaluation no longer discriminates at the
frontier.
The same conclusion arises from detection: the largest recent AI-agent evaluation on real CVEs surfaced 18 incomplete patches across 15 projects during scoring alone~\cite{wang2025cybergym}, previewing the empirical picture we consider next.

These empirical limits generalize. On CyberGym's $1,507$-instance benchmark spanning $188$ real projects, frontier agents reproduce the target vulnerability in only $17-22$\% of cases, surfacing the $18$ incomplete patches referenced above~\cite{wang2025cybergym}.
Detection benchmarks show similar erosion once measurement flaws are corrected: after removing mislabeled examples and training-set leakage, top LLMs drop from 68\% to 3\% F1-score, and GPT-4 with chain-of-thought fails to distinguish
vulnerable functions from their patched versions in paired comparisons~\cite{ding2025vulnerability}. 
A structural cause underlies both effects: recent measurement of the maximum effective context window finds effective capacity falling short of the advertised window by as much as 99\%, with a few models failing at as few as 100 tokens of context and most degrading severely by 1{,}000 tokens~\cite{paulsen2026contextwindow}. 

\sol takes this picture as its design premise but adds artifacts these general-purpose systems do not provide. 
The benchmark and evaluation works~\cite{wang2025cybergym, ding2025vulnerability,
paulsen2026contextwindow, JitVul2025benchmarking} identify these limits but propose no method to overcome them.
Kumushi~\cite{wang2026kumushi} is the exception because it proposes a repair method. \sol{} applies its symptom-fix taxonomy retrospectively (§~\ref{sec:framework}).
The primary contribution is the integration of four components into a 5G-specific audit procedure. Two components are domain-specific: a 17-entry cellular-core anti-pattern catalog and 3GPP-procedure-driven reachability triage. Concentric validation orders effort by testbed cost. Patch-review applies Kumushi's taxonomy to shipped CVE fixes rather than newly generated patches.

\section{\sol Framework}
\label{sec:framework}

\begin{figure*}[!t]
\centering
\begin{tikzpicture}[
    node distance=0.4cm and 0.4cm,
    stepbox/.style={draw, rectangle, minimum width=1.6cm, minimum height=1.1cm, align=center, font=\small, thick},
    aibox/.style={stepbox, fill=green!15},
    humanbox/.style={stepbox, fill=blue!15},
    mixedbox/.style={stepbox, fill=orange!25},
    scaffold/.style={draw, dashed, rectangle, rounded corners=3pt, inner sep=6pt},
    axis/.style={-Latex, thick},
]
\node[aibox] (s0) {\textbf{0}\\[-2pt]\scriptsize Baseline\\[-3pt]\scriptsize tools};
\node[humanbox, right=of s0] (s1) {\textbf{1}\\[-2pt]\scriptsize Select\\[-3pt]\scriptsize patterns};
\node[aibox, right=of s1] (s2) {\textbf{2}\\[-2pt]\scriptsize Enumerate\\[-3pt]\scriptsize sites};
\node[mixedbox, right=of s2] (s3) {\textbf{3}\\[-2pt]\scriptsize Inspect\\[-3pt]\scriptsize context};
\node[humanbox, right=of s3] (s4) {\textbf{4}\\[-2pt]\scriptsize Establish\\[-3pt]\scriptsize reachability};
\node[mixedbox, right=of s4] (s5) {\textbf{5}\\[-2pt]\scriptsize Validate\\[-3pt]\scriptsize impact};
\node[humanbox, right=of s5] (s6) {\textbf{6}\\[-2pt]\scriptsize Select\\[-3pt]\scriptsize fix layer};
\foreach \a/\b in {s0/s1, s1/s2, s2/s3, s3/s4, s4/s5, s5/s6} {
    \draw[axis] (\a) -- (\b);
}
\node[scaffold, fit=(s1)(s2), label={[font=\small\itshape, yshift=-2pt]above:{Anti-pattern catalog}}] {};
\node[scaffold, fit=(s3)(s4), label={[font=\small\itshape, yshift=-2pt]above:{Critical-path triage}}] {};
\node[scaffold, fit=(s5)(s6), label={[font=\small\itshape, yshift=-2pt]above:{Concentric validation}}] {};
\node[below=0.4cm of s3, font=\small\itshape, align=center, text width=4cm] (patch) {Published CVE patch};
\draw[axis] (patch.east) -| (s5.south);
\end{tikzpicture}
\caption{The \sol procedure: seven steps grouped by function, with published CVE patches entering at Step 5. Step boxes are color-coded by execution role: \colorbox{green!15}{\strut AI-automatable}, \colorbox{orange!25}{\strut mixed (AI + human)}, \colorbox{blue!15}{\strut human-required}.}
\Description{Diagram of the seven-step \sol procedure. A published CVE patch enters at Step 5. Step colors indicate AI-automatable, mixed, or human-required work.}
\label{fig:drishti}
\end{figure*}
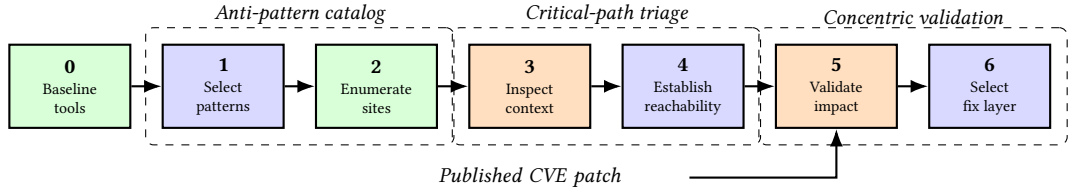

\subsection{Overview}

Figure~\ref{fig:drishti} shows \sol at a glance. \sol is a seven-step procedure that the audit operator runs against a target module. 
The steps are (0) run automated linters, (1) pick the anti-pattern set in scope, (2) enumerate candidate sites from the linter output combined with the catalog's grep recipes, (3) contextual check by manual or AI-led reading, (4) reachability cross-reference against an attacker-model taxonomy, (5) validate exploitability through concentric validation, and (6) choose the fix layer. 
The procedure has a specific origin. 
An early audit of free5GC's NAS library wrote up and patched a single-site defect. 
The patch did not restore baseline behavior under stress. 
Investigation surfaced 45 sites of the same anti-pattern in the same library, and we withdrew the single-site framing. 
The procedure codifies the missing ``is this a class?'' step.

The catalog-driven passes also record candidate attrition. The NAS AP1 pass produced 89 raw matches and 45 confirmed in-scope sites. Eight information-element types were reachable before authentication. The NGAP AP1 pass found one shared decoder site used by 304 callers. Appendix~\ref{app:audit-accounting} reports the negative, sampled, and deferred categories.

\sol runs AI-led, with human work focused on catalog curation, triage criteria, 3GPP reasoning, lab construction, disclosure decisions, and final cross-checks.
Each is amortized across many audits. 
The AI does the per-candidate bulk: source-file reading at depth, catalog-driven enumeration, contextual checks, unit-test harness drafting, and patch-diff re-reading. 
The audit work ran on Anthropic's Claude Opus models, spanning versions 4.7 and 4.8 across sessions.

\sol is cost-aware throughout: each step spends the next step's resources only on candidates that survive cheaper upstream filters. Linters are free per invocation but blind to the highest-impact decoder anti-patterns; the catalog's grep recipes bound the AI's reading scope, and so its token budget, to candidate sites; and concentric validation advances only on confirmed downstream effect, which keeps lab time off candidates that fail a unit test.

\subsection{Anti-pattern catalog}

The catalog is a structured list of recurring vulnerability-producing code patterns in protocol-decoder, network-handler, and IPC code. It holds 17 entries with the AP1 through AP17 identifiers and two protocol-specific sections populated through audit passes against free5GC NAS (TS 24.501) and free5GC NGAP plus the underlying ASN.1-PER decoder (TS 38.413, ITU-T X.691 \cite{itu2021x691}). Per-entry structure covers a description, the idiom in each in-scope language, a grep recipe for mechanical candidate enumeration, the runtime symptom, false-positive considerations, common fix layers (per-call-site, per-type, per-message, per-NF), the CWE class the entry maps to, and known instances from prior audits.

The catalog indexes patterns. It is not the authority a finding cites. Findings cite external standards: CWE for the weakness class, SEI CERT C and MISRA C for C surfaces such as Open5GS, gosec and staticcheck rule identifiers for Go surfaces, and the relevant 3GPP TS clause for protocol-conformance bugs \cite{3gpp2025ts24501,3gpp2025ts38413,3gpp2025ts33501}.

A worked example: AP1, \emph{unbounded allocation on attacker controlled length}, captures sites where a buffer is allocated with a size that comes from network input without an upper-bound check. In Go, the idiom is \texttt{make({[}{]}T,\ N)}, where \texttt{N} is read from peer input. The runtime symptom is memory amplification under sustained attack: per-request RAM allocation scales with declared \texttt{N}, and garbage-collector pressure dominates request handling before any out-of-memory kill. Fix layers run from per-call-site length checks through per-type bounds in the allocator method to per-message dispatcher caps at the NF entry. The CWE mapping is CWE-789. Known AP1 instances to date include 45 sites in the free5GC NAS library (with 8 reachable from a pre-authentication UE peer) and 1 site in the ASN.1-PER decoder reachable from 304 NGAP decoder call sites via the encoding's \texttt{sizeUB:65535} tag.

The catalog exists to cover what linters miss. We compared the documented rule coverage of \texttt{gosec}, \texttt{staticcheck}, and \mbox{\texttt{golangci-lint}} with the catalog. None has a rule for the highest-impact decoder anti-patterns: memory amplification (AP1), CPU exhaustion (AP2), stack exhaustion (AP3), and unbounded state growth (AP6). Detecting them calls for analyses these tools do not carry by default: taint tracking for AP1 and AP2, an AST query for AP3, and handler-state review for AP6. Where a linter has no rule, the catalog's grep recipe is the enumerator.

\subsection{Critical-path triage}

Before any audit work, we rank attack surfaces and drop code paths already covered by published CVEs to avoid rediscovery cost. 
Each finding in this paper falls into one of two origin classes. 
A novel discovery is a defect found by targeted source-reading or a catalog match on code with no prior published CVE. A patch-review starts from the patched region of a published CVE and re-reads it from the original attacker's position. The patch can introduce the defect or leave the intended fix as unreachable code.

We rank the surviving surface by 3GPP-procedure impact. UE registration, NF discovery, session establishment, and authentication exchange carry direct operator-observable consequence on compromise and rank above logging, metrics, and configuration-fetch procedures. Pre-authentication paths rank above post-au\-thentication paths because they do not require credential compromise. Cell-area or serving-area blast radius defects rank above single-UE-session defects. Control-plane defects with cascade effects on registered NFs rank above isolated-NF defects.

Reachability cross-reference asks, for each candidate, which attacker can deliver the input and in what state. The two pre-authentication classes from Section 2 --- a rogue UE sending NAS over the air, and a rogue gNB sending NGAP over an SCTP association --- are the ones we audit here. We tag each candidate with the class that reaches it and the 3GPP procedure that carries the input, which gives an ordered list of code regions to read closely.

\subsection{Concentric validation}

We use \emph{concentric} as expanding scope, not layered defense: the bug sits at the center, and validation expands outward through four stages, each adding context and validating a stronger claim than the last at steeply higher cost (Table~\ref{tab:stages}).

\begin{table*}[t]
\caption{Concentric validation stages.}
\label{tab:stages}
\begin{tabularx}{\textwidth}{@{}cXXc@{}}
\toprule
Stage & Context & Claim validated & AI-automatable? \\
\midrule
1 & Suspect function, unit test & The bug exists & Yes \\
2 & Single NF, crafted protocol message & The bug crashes the NF & Partial \\
3 & Full 5G core, realistic UE/gNB & The bug breaks the deployed system & \textbf{No} \\
4 & Operator-relevant measurement & The bug has measured blast radius & \textbf{No} \\
\bottomrule
\end{tabularx}
\end{table*}

We read ``AI-automatable'' as the current capability of large language models and adjacent tooling. Stage~2 is only partial because input crafting and symptom interpretation benefit from human cross-check, and Stages~3 and~4 need lab infrastructure with realistic UE/gNB traffic and 3GPP semantic interpretation of system-level symptoms.

Stage 1 isolates the suspect code into a unit-test harness and exercises it with the input that triggered the audit flag. The bug either reproduces or it does not. This stage cheaply discards false positives and produces a per-request resource signature.

Stage 2 brings up exactly one network function. We deliver the candidate input as a crafted protocol message and record crashes, hangs, memory pressure, and CPU saturation. This stage establishes that the defect transits through the protocol decoder and produces a network-function-level effect.

Stage 3 runs the full 5G core under realistic UE and gNB traffic. We provision subscribers in a UDM, and UE attach and authentication flows complete. We inject the candidate input from the appropriate adversary position, using a UERANSIM-driven rogue UE for NAS-path defects and a custom SCTP client for NGAP-path defects. Stage 3 is where 3GPP semantic knowledge is decisive. The message has to be valid enough at every protocol layer to reach the defect. The observation has to distinguish a stuck UE state machine from an outright network-function crash.

Stage 4 instruments the deployment to measure operator-relevant impact: time to impact, scope, recovery semantics, and configuration sensitivity. The outputs are the numbers an operator would use to decide procurement, mitigation, and patch priority.

When validation is complete, the audit recommends a fix at the layer the anti-pattern admits: a per-call-site check, a per-type bound in a shared allocator, a per-message cap at the dispatcher, or a per-NF cgroup or rate limit.

\subsection{patch-review}

The patch-review component applies the same validation to a different starting point. Instead of starting from a module and a catalog pattern, it starts from a published CVE patch and re-reads the patched code from the attacker's position.
The defect is confirmed if either the original input still produces an attacker-relevant outcome, or a related input the patch was meant to reject is accepted because the new rejection code is unreachable.
We label confirmed patch defects with Kumushi's nine-category symptom-fix taxonomy~\cite{wang2026kumushi}. Kumushi developed the taxonomy for generated patches. We apply it retrospectively to shipped CVE fixes.
The later validation stages are unchanged. 
The catalog and triage are not needed here, because the patched code region is already the entry point.


\section{Case studies}
\label{sec:findings}
We present three case studies drawn from the two audited cores. 
Section~\ref{sec:case-ngap} applies the full \sol procedure to decoder-side amplification in free5GC's NGAP.
Section~\ref{sec:case-nrf} demonstrates targeted source-reading when no catalog pattern matches. 
Section~\ref{sec:case-patch} shows patch-review applied to a published CVE fix.

\subsection{Running example: free5GC NGAP amplification}
\label{sec:case-ngap}
We apply each step of the \sol procedure to the free5GC NGAP amplification finding.
Step 0 ran gosec and staticcheck against the free5GC NGAP and ASN.1-PER libraries. 
gosec returned 35 hits, all G115 integer-narrowing warnings of the type AP12 documents as dominated by spec-bounded narrowings in 3GPP IE setters. 
staticcheck returned no security-relevant issues. 
The libraries appeared clean to off-the-shelf static analysis. 
Step 1 picked AP1, AP2, AP3, AP5, AP8, AP12, and AP15 as the in-scope set. 
Step 2 enumerated. 
AP1 produced exactly one decoder-side site: a single \texttt{reflect.MakeSlice} call inside the SEQUENCE-OF decoder, reachable from 304 ngapType callers via the ASN.1 \texttt{sizeUB:65535} schema tag. 
Step 3 contextually checked the site by manual reading of the 980-line decoder. 
The SEQUENCE-OF decoder pre-allocates before reading element data. OCTET STRING and BIT STRING decoders are well-defended with byte-cursor gates. 
Step 4 cross-referenced reachability. 
NGSetupRequest, the first NGAP message after SCTP association establishment, decodes through this path with zero authentication state. 
The attacker class is \emph{peer gNB pre-handshake}, requiring only an SCTP association to the AMF in deployments without N2 IPsec.

Step 5 ran concentric validation. Stage 1 measured 4 MB of heap allocated by the decoder per 2-byte malformed input, an amplification factor exceeding two million. Stage 2 brought up an AMF with \texttt{MemoryMax=256M} and 48 concurrent workers, and the kernel OOM-killed the AMF process in 6.2 seconds. A single-NF OOM-kill is decisive for a memory-exhaustion bug, so validation stopped there. Appendix A characterizes how the time-to-OOM varies with the cgroup limit and worker count. Step 6 identified the fix layer. A single bound on the element count in the ASN.1-PER decoder would close all 304 NGAP call sites at once. We recommend pairing it with a per-NGAP dispatcher cap at the AMF for deployment-side defense-in-depth.

\subsection{Open5GS NRF NULL-dereference, targeted source-reading}
\label{sec:case-nrf}
This case shows \sol finding a defect without a catalog match. The NRF SBI multipart parser is reachable before authentication because the defect occurs before request dispatch. AI-led source reading found a NULL field dereference in the header-value path. Stage 1 reproduced the defect in a unit test, and Stage 2 crashed one NRF with the first malformed message. Stage 3 confirmed that the NRF crash sent all seven registered network functions into exception and retry states. About 90 seconds later, the SMF logged \texttt{No\ UPF\ available} and de-associated its PFCP connection to the UPF. This state prevents new PDU-session establishment until recovery, although the test did not attempt a new session after PFCP loss. In a similarly configured private 5G deployment, this failure would interrupt devices that need a new or recovered session. We did not test a physical-process consequence. The maintainer fixed the defect within 15 days, and we requested a CVE from MITRE.

\subsection{Defective patch on CVE-2025-69248}
\label{sec:case-patch}
This case shows the patch-review component. The published v1.2.3 patch on free5GC's NAS library closes the original AMF panic correctly but adds a defense-in-depth validation block that is unreachable on the pre-authentication path due to two coding errors: a read against a zero-allocated buffer before the wire read, and a mask width that misclassifies the identity-type field against TS 24.501 §9.11.3.4 (cf.~App.~\ref{app:patch-details}).
Stage 1 used seven unit sub-tests, each constructing a \texttt{RegistrationRequest} with a violating \texttt{MobileIdentity5GS} value. 
All seven pass. 
The decoder accepts every violation the validation block was intended to reject. 
The dead block does not crash the AMF on its own, so the finding needs no live-stack reproduction. 
The unit tests alone show the intended hardening is inert. 
The defense-in-depth layer is eliminated for a class of pre-authentication inputs. 
A CVE patch that adds defense-in-depth requires an audit step beyond verifying that the original input no longer reaches the original sink. 
We disclosed it through the project's private vulnerability-reporting channel.

In Kumushi's taxonomy~\cite{wang2026kumushi}, error~1 is a wrong-location repair and error~2 is an ineffective guard. The two categories occur in one 30-line patch block.

\section{Discussion}
\label{sec:discussion}

CoreCrisis~\cite{dong2025corecrisis} targets the same pre-authentication NAS/N1 surface as our NGAP amplification finding, but through stateful FSM fuzzing of a black-box implementation. 
The two methods discover different defect classes: FSM fuzzing surfaces state-transition bugs the model can express, while source reading against a domain catalog surfaces decoder-side allocation defects reachable before the state machine engages. 
FivGeeFuzz~\cite{chen2025cross} occupies the opposite surface, fuzzing SBIs post-authentication with grammars derived from 3GPP OpenAPI specs. 
Across the three findings the four costs were exercised unevenly: verification took three different routes, reachability was decisive for all three because each was pre-authentication, and impact justified the staged design, with the NGAP finding stopping at Stage~2 and the NRF finding needing Stage~3.

Our results are illustrative rather than statistical. The three generation routes have different denominators, so one false-positive rate would be misleading. We examined only open-source implementations and did not run a controlled comparison against another audit system. The CPS/IoT experiment measures 5G service failure, not a physical-process consequence. Future work should examine post-authentication surfaces and compare independent implementations, which separates implementation defects from hazards in the protocol or encoding.

\section{Conclusion}
Generating vulnerability candidates in 5G core code is increasingly within reach of AI methods. 
What remains is turning those candidates into claims about real operator impact, and spending audit effort only where evidence warrants it. 
\sol pairs a cellular-core anti-pattern catalog and critical-path triage with concentric validation and patch-review, so that each of the four validation costs is paid only where cheaper evidence says it is warranted. 
Applied to Open5GS and free5GC, it produced three findings, including a 2-byte pre-authentication input that OOM-kills the free5GC AMF in 6.2~seconds and a shipped CVE patch whose defense-in-depth check is dead code. 
Comparing independent open-source cores and applying patch-review beyond 5G are the natural next steps.


\section*{Acknowledgement}
This research/project is supported by the National Research Foundation, Singapore, and the Cyber Security Agency of Singapore under the National Cybersecurity R\&D Programme and the CyberSG R\&D Programme Office (Award CRPO-GC2-ASTAR-001).
Any opinions, findings, conclusions, or recommendations expressed in these materials are those of the author(s) and do not reflect the views of the National Research Foundation, Singapore, the Cyber Security Agency of Singapore, or the CyberSG R\&D Programme Office.

\section*{Ethics and Disclosure}
\label{sec:ethics}
All experiments ran against open-source 5G core implementations in an isolated laboratory, never against a production network or a third party's deployment. 
We disclosed every finding to the affected project before submitting this paper: the Open5GS NRF dereference through the maintainer's coordinated channel, with a CVE requested from MITRE, and the free5GC NAS and NGAP findings through the project's private reporting process. 
We follow standard disclosure timelines and shared reproduction details to support remediation.

\bibliographystyle{acm_bibtex_format}
\bibliography{main}

\appendix
\section{Appendix}
\label{sec:appendix}
\subsection{ASN.1-PER amplification: detailed measurements}
\label{app:ngap-details}
The vulnerable decoder is \texttt{parseSequenceOf} in the \texttt{free5gc/aper} module. 
ASN.1 Aligned Packed Encoding Rules use a length-prefixed encoding for \texttt{SEQUENCE\ OF} and similar container types. 
The decoder reads the wire-supplied element count first and passes it to \texttt{reflect.MakeSlice(sliceType,\ intNumElements,\ intNumElements)} before any element is parsed (the repeated argument sets slice length and capacity to the same wire-supplied count). 
No upper bound is applied to the element count, despite the schema tag \texttt{sizeUB:65535}. 
Across the NGAP type set, 304 fields are tagged this way, and every standard NGAP message container exposes a \texttt{parseSequenceOf} call on this path.

Under unit-test isolation, a 2-byte malformed input (\texttt{0xFF\ 0xFF} patched into the element-count field of an otherwise well-formed \texttt{NGSetupRequest}) drives the decoder to allocate 4,200,576 bytes (4.01 MB) of heap before parsing any element. 
Per-element struct size for the target IE type is 64 bytes. 
The amplification factor exceeds two million.

Time to OOM is host-dependent. 
Table~\ref{tab:appendix-oom} summarizes the original three deployment configurations and one cross-host revalidation.
Without a cgroup memory cap, AMF resident set size grows to $447$~MB ($8.4\times$ baseline), and virtual memory grows to $626$~MB. Resident memory recovers after the attack stops, but the Go heap retains the 626~MB reservation.
Under a cgroup cap, SBI responsiveness degrades during the attack, dropping to $71$\% in the \texttt{MemoryMax=512M}, 24-worker run before the OOM-kill.

\begin{table}[ht]
\caption{AMF time-to-OOM under sustained malformed NGSetupRequest traffic. The original runs used an aarch64 host. The final row is an x86\_64 revalidation.}
\label{tab:appendix-oom}
\begin{tabularx}{\columnwidth}{@{}Xl@{}}
\toprule
Configuration & Time to OOM-kill \\
\midrule
no \texttt{MemoryMax} cap (aarch64) & no OOM (RSS 447 MB) \\
\texttt{MemoryMax=512M}, 24 workers (aarch64) & 42.6 s \\
\texttt{MemoryMax=256M}, 48 workers (aarch64) & 6.2 s \\
\texttt{MemoryMax=256M}, 48 workers (x86\_64) & about 31 s \\
\bottomrule
\end{tabularx}
\end{table}

The defect is sensitivity-tunable but not avoidable through process limits alone. 
The upstream fix has to bound the decoder's allocation. 
The wider hazard sits above this codebase.
ASN.1-PER encodes length determinants for these constructs~\cite{itu2021x691}.
An implementation that allocates directly from an unbounded wire-supplied count can therefore exhibit the same defect class, independently of this codebase.

\subsection{CVE-2025-69248 patch defect: line-by-line analysis}
\label{app:patch-details}
CVE-2025-69248~\cite{nvd2025cve} is a published vulnerability in the \texttt{free5gc/nas} module. 
The vulnerability is an AMF panic in the \texttt{MobileIdentity5GS} decoder. A malformed value in a pre-authentication NAS \texttt{RegistrationRequest} triggers it.
The v1.2.3 patch contains two changes. 
The first change closes the original panic path, which we confirmed by reading the patch. 
The second change adds a defense-in-depth validation block intended to reject identity values that violate the per-type length constraints from TS 24.501 \S9.11.3.4. 
The validation block contains two coding errors that, in combination, make the entire block unreachable.

\paragraph{Coding error 1: zero-allocated buffer read.} 
At \texttt{nasMessage/NAS\_RegistrationRequest.go:321}, \texttt{SetLen()} is called. Internally this executes \texttt{Buffer\ =\ make({[}{]}uint8,\ Len)}, which produces a zero-allocated Go slice. At line 322 a length-guard executes (\texttt{if\ a.MobileIdentity5GS.Len\ \textgreater{}\ 0}). At line 323 the validation block reads \texttt{a.MobileIdentity5GS.Buffer{[}0{]}\ \&\ 0x0F} to extract the identity type. The actual wire bytes are not copied into \texttt{Buffer} until line 353, via \texttt{binary.Read(buffer,\ binary.BigEndian,\ a.MobileIdentity5GS.Buffer)}. The line-323 read therefore always returns \texttt{0x00}. The validation switch always takes the \texttt{MobileIdentity5GSTypeNoIdentity} case (type value 0). Every other case, including SUCI, 5G-GUTI, IMEI, 5G-S-TMSI, and the default rejection branch, is dead code.

\paragraph{Coding error 2: wrong mask width.} 
The mask \texttt{0x0F} selects bits 1 through 4 of the first identity byte. Per TS 24.501 §9.11.3.4, the identity-type field occupies bits 1 through 3 only (mask \texttt{0x07}). Bit 4 is an independent odd/even indicator. Even if error 1 were fixed, the wrong mask would mis-classify the four odd-encoded identity types (\texttt{0x09}, \texttt{0x0B}, \texttt{0x0D}, \texttt{0x0F}) and would still fail to reject undefined identity types in the way the spec requires.

\paragraph{Unit-test confirmation.} 
We constructed seven sub-tests against \texttt{free5gc/nas@v1.2.3}. 
Each constructs a \texttt{RegistrationRequest} byte sequence with a violating \texttt{MobileIdentity5GS} value: undersized GUTI, undersized IMEI, undersized 5G-S-TMSI, undersized SUCI, undefined identity types (\texttt{0x06}, \texttt{0x07}), and an oversized length. 
All seven sub-tests pass. 
The decoder accepts every violation the validation block was intended to reject.

\paragraph{Why no live-stack reproduction was needed.} 
The dead validation block does not crash the AMF on its own. 
The original CVE-2025-69248 panic is closed by the patch's other change, which we confirmed by reading the patch. 
What does not take effect is the intended hardening. 
The validation block is dead in the running AMF, and any pre-authentication UE submitting a malformed \texttt{MobileIdentity5GS} is silently accepted by the decoder. 
The practical impact is the elimination of the intended defense-in-depth layer for a class of pre-authentication inputs whose individual exploitability remains an open question for downstream code.

\subsection{Audit accounting}
\label{app:audit-accounting}
Table~\ref{tab:audit-accounting} reports candidate counts from the recorded audit trails. ``Raw'' is the number of mechanical matches before scope exclusions. ``In scope'' is the number retained for contextual review. Parentheses show the reviewed sample when review was incomplete. ``Confirmed'' means that source review established the code pattern, not that exploitability was proven. The final column reports reachability and runtime validation. Targeted source reading and patch-review have no comparable raw-candidate denominator. We therefore do not combine the routes into one false-positive rate.

\begin{table*}[ht]
\caption{Candidate accounting. A dash means that the later stage was not applicable or was not performed.}
\label{tab:audit-accounting}
\begin{tabularx}{\textwidth}{@{}lrrrX@{}}
\toprule
Audit cell & Raw & In scope & Confirmed & Reachability and validation \\
\midrule
NAS AP1 & 89 & 45 & 45 & 8 pre-authentication IE types, one representative cell validated \\
NAS AP2 & 10 & 3 & 0 & -- \\
NAS AP5 & 13 & 2 & 0 & -- \\
NAS AP8 & 983 & deferred & -- & Search was too noisy without a decoder filter \\
NAS AP12 & 58 & 58 (5 reviewed) & 0 in sample & Exhaustive review deferred \\
NGAP/APER AP1 & 1 shared site & 1 & 1 & Pre-handshake, unit test and live AMF \\
NGAP/APER AP2 & 1 & 1 & 0 distinct & Same site and effect as AP1 \\
NGAP/APER AP12 & 35 & 35 (3 reviewed) & 0 in sample & Exhaustive review deferred \\
NGAP/APER AP3 & 1 & 1 & 0 & ASN.1 grammar bounds recursion depth \\
Open5GS NRF & -- & targeted read & 1 & Pre-authentication, unit, single-NF, and full-core validation \\
CVE patch & 1 block & 1 & 2 errors & Pre-authentication path, seven unit sub-tests \\
\bottomrule
\end{tabularx}
\end{table*}

\subsection{Experimental setup}
\label{app:experimental-setup}
Table~\ref{tab:experimental-setup} summarizes the setup for each reported runtime result. Early audit records did not capture operator time consistently. We therefore report experiment duration and stopping conditions, not end-to-end audit time.

\begin{table*}[ht]
\caption{Experimental setup for the three case studies.}
\label{tab:experimental-setup}
\begin{tabularx}{\textwidth}{@{}lXXX@{}}
\toprule
Case & Target & Input and attacker position & Environment and stopping condition \\
\midrule
Open5GS NRF & Open5GS v2.7.5 & 47-byte HTTP/2 multipart request from an SBI peer, then from a UE data session in a lab without CP/UP segregation & Parser harness, single NRF, and Docker full core. Stop after the first crash, retry cascade, and PFCP loss at about 90 s \\
free5GC NGAP & free5GC main, \texttt{ngap} v1.1.3, \texttt{aper} v1.1.1 & Stage 1 used a 2-byte decoder input. Stage 2 sent malformed \mbox{NGSetupRequest} messages from a peer gNodeB & Original runs used an aarch64 host with 8--48 workers and no cap, 512 MB, or 256 MB limits. An x86\_64 revalidation used 48 workers and 256 MB \\
CVE-2025-69248 patch & \texttt{free5gc/nas} v1.2.3 & Seven invalid \mbox{RegistrationRequest} inputs from the pre-authentication UE path & Go unit tests. Acceptance of an input that the new block should reject confirms the defect \\
\bottomrule
\end{tabularx}
\end{table*}
\end{document}